\documentclass[aps,prl,twocolumn]{revtex4}

\usepackage{amsthm}
\usepackage{epsfig}
\newcommand{\beq}{\begin{equation}}
\newcommand{\eeq}{\end{equation}}
\newcommand{\beqa}{\begin{eqnarray}}
\newcommand{\eeqa}{\end{eqnarray}}
\newcommand{\beqar}{\begin{eqnarray*}}
\newcommand{\eeqar}{\end{eqnarray*}}
\newcommand{\tr}{{\rm tr}}

\newcommand{\bei}{\begin{itemize}}
\newcommand{\eei}{\end{itemize}}
\newcommand{\bee}{\begin{enumerate}}
\newcommand{\eee}{\end{enumerate}}
\def\be{\begin{equation}}
\def\ee{\end{equation}}
\def\ben{\begin{eqnarray}}
\def\een{\end{eqnarray}}
\def\eea{\end{array}}
\def\bea{\begin{array}}
\def \s {\,\,\,\,}

\def \ra {\rangle}

\def\rab{\rho_{AB}}
\def\vr{\varrho}

\newcommand{\proj}[1]{\ket{#1}\bra{#1}}
\newcommand{\bra}[1]{\langle #1 |}
\newcommand{\ket}[1]{| #1 \rangle}

\def \E {E_r}

\def \K {K_D}
\def \D {E_D}
\def \singlet {\psi^{+}}

\def \X {X}
\def \upl {\sigma}
\def \lowr {\sigma'}

\def\textbf#1{{\bf #1}}
\def\blacksquare{\vrule height 4pt width 3pt depth2pt}

\def\duzomniejsze{<\kern-.7mm<}
\def\duzowieksze{>\kern-.7mm>}

\def\textbf#1{{\bf #1}}

\def\tr{{\rm Tr}}

\def\>{\rangle}
\def\<{\langle}

\def\ot{\otimes}
\def\pb{\gamma}

\def\sep{sep}

\def\Ps{P_{sym}}
\def\Pa{P_{as}}

\def\Logneg{E_N}
\def\erinf{E_r^\infty}

\begin{document}

\title{Secure key from bound entanglement}


\author{Karol Horodecki$^{(1)}$,
Micha\l{} Horodecki$^{(2)}$,
Pawe\l{} Horodecki$^{(3)}$,
Jonathan Oppenheim$^{(2)(4)(5)}$
}
\affiliation{$^{(1)}$Department of Mathematics, Physics and Computer Science,
  University of Gda\'nsk, Poland}
\affiliation{$^{(2)}$Institute of Theoretical Physics and Astrophysics,
University of Gda\'nsk, Poland}
\affiliation{$^{(3)}$Faculty of Applied Physics and Mathematics,
Technical University of Gda\'nsk, 80--952 Gda\'nsk, Poland}
\affiliation{$^{(4)}$
Dept. of Applied Mathematics and Theoretical Physics, University of
Cambridge, Cambridge, U.K.}
\affiliation{$^{(5)}$
Racah Institute of Theoretical Physics, 
Hebrew University of Jerusalem, Givat Ram, Jerusalem 91904, Israel}


\affiliation{}


\date{September 11th, 2003}
\begin{abstract}
We characterize the set of shared quantum states which contain
a cryptographically private key.
This allows us to recast the theory of privacy
as a paradigm closely related to that used in entanglement manipulation.
It is shown that one can distill an arbitrarily secure key
from bound entangled states.  
There are also states which have less
distillable private key than the entanglement cost of the state.  
In general the
amount of distillable key is bounded from above by the relative
entropy of entanglement.  
Relationships between distillability and distinguishability are found
for a class of states which have Bell states correlated to separable 
hiding states. We also describe a technique for 
finding states exhibiting irreversibility in entanglement distillation. 

\end{abstract}

\pacs{}
\keywords{}


\maketitle


Recently, strong connections have been emerging between the amount of 
pure entanglement $E_D$ and the private key $K_D$ one can distill from a shared 
quantum state.
For example, the security of key generation in BB84 \cite{bb84} and B92 \cite{b92}
can be proven by showing its equivalence with
entanglement distillation of singlets \cite{shor-preskill}\cite{b92-secure}.  
These proofs had their origin in the idea of {\it quantum privacy amplification}
\cite{QPA} where two parties (Alice and Bob) distill pure quantum entanglement
until the quantum correlations are completely disentangled with an
eavesdropper (Eve). Those correlations were represented by singlet states
and were subsequently measured to 
obtain a classical private key to which Eve had no access.
Very recently, the hashing inequality
\cite{BDSW1996,HHH-hashing} was proven \cite{DevetakWinter-hash} by showing the 
equivalence between
certain distillation protocols and
one way secret key distillation. 

An apparent equivalence between
bound entangled states (states which require entanglement to create,
but from which no pure entanglement can be distilled) and classical distributions which
can not be turned into a key was conjectured in
\cite{boundinfo-gw}.  
Additionally, using techniques developed in entanglement
theory, a gap similar to the one between entanglement cost and
distillable entanglement was shown to exist classically for private keys
\cite{renner-wolf-gap}.  It has also been shown that for two qubits,
a state is one copy distillable iff it is cryptographically secure \footnote{In the case
where the eavesdropper measures her states before privacy amplification} 
\cite{acin-gisin} (c.f. \cite{gisin-wolf-99,bruss-tomo-crypto}), and there are basic laws which govern the
interplay of key generation in terms of sent quantum states \cite{oh-recycling}.

In fact, the original papers on entanglement distillation \cite{BDSW1996}
used protocols which were derived from existed protocols for distilling
privacy from classical probability distributions.  Indeed, formal analogies
between entanglement and secrecy exist \cite{collins-popescu}.
The evidence to date strongly supports the widely
held belief that privacy and entanglement distillation are strictly equivalent
-- that
one can get a private key from a quantum state if and only if 
entanglement distillation
is possible.


Surprisingly, this is not the case - we will introduce 
a class of {\it bound entangled states} (no pure entanglement can be distilled from them),
from which one can distill a private key. 
Examples of states that have one bit of perfect private key and
at the some time arbitrarily small distillable entanglement 
are also provided.

Clearly, one always has $\K\geq\D$ since one can 
always distill singlets
from a state, and then use these singlets to generate a private key
\cite{ekert-qkd}.
Here, we prove that one can also have the strict inequality $\K>\D$, 
which sometimes holds even if $\D=0$. 
We will also prove that the private key is generally bounded from above
by the relative entropy of entanglement $\E$ \cite{PlenioVedral1998} 
(regularized).  This will be
sufficient to prove that one can have $\K<E_c$ where $E_c$ is the
entanglement cost (the number of singlets required to prepare a state under
LOCC).  
This enables one to easily find states for which $\D < \E $.
In the present paper we will state some of the results and present the full
proofs in detail elsewhere \cite{keyhuge}.

We will first introduce a wide class of states which are the
most general {\it private states} in the sense that one
can  produce one bit of secure key from them even though an
eavesdropper might hold the purification of the state. One can think
of these states as being the equivalent of the singlet for key distillation.
This will allow us to recast all protocols of key distillation (classical or otherwise)
in terms of distillation of private states using the distant labs paradigm
used in entanglement theory i.e local operations and classical communication (LOCC).
Next we show that these states can have arbitrary
little distillable entanglement while still retaining one bit of
private key.  We can relate this to the problem of distinguishability
of states under LOCC.  We then exhibit a bound entangled state from
which a private key can be distilled.  
We then prove that $\K\leq\E$ and discuss the consequences.

Let us now introduce private states i.e. $\gamma_{ABA'B'}$ 
where systems $AB$ are both $m$-qubits,  
and  measurement of $AB$ in the computational basis gives  
$m$ bits of perfect key. Systems $AA'$  ($BB'$) are held by Alice (Bob). 
We assume the usual scenario - that any part of the state which is not 
with Alice and Bob might be with an eavesdropper 
Eve.  Thus Eve holds the purification of this state. 
We will now provide their unique form.  We first consider 
perfect security.

{\it Theorem 1. A  state is private in the above sense  
iff it is of the following form
\beq
\pb_m=U|\singlet_{2^m}\>_{AB}\<\singlet_{2^m}|\ot \vr_{A'B'}U^\dagger
\label{eq:pstate}
\eeq
where $\ket{\singlet_{d}}=\sum_{i=1}^d |ii\ra$ 
and $\vr_{A'B'}$ is an arbitrary state on $A'$,$B'$.   
$U$ is an arbitrary unitary controlled in the computational basis 
\beq
U=\sum_{i,j=1}^{2^m} |ij\>_{AB}\<ij|\otimes U_{ij}^{A'B'} \s .
\label{eq:u}
\eeq }

We will call the operation (\ref{eq:u}) "twisting" (note that only $U_{ii}^{A'B'}$  matter here, 
yet it will be useful to consider general twisting later).  

{\it Proof.}
We will prove for $m=1$ (for higher $m$, the proof is analogous).
Start with an arbitrary state held by Alice and Bob, $\rho_{AA'BB'}$,
and include its purification 
to write the total state in
the decomposition
\beqa
&&  \Psi_{ABA'B',E}=
a|00\rangle_{AB}|\Psi_{00}\rangle_{A'B'E} + 
b|01\rangle_{AB}|\Psi_{01}\rangle_{A'B'E} \nonumber\\
&&+c|10\rangle_{AB}|\Psi_{10}\rangle_{A'B'E} +
d|11\rangle_{AB}|\Psi_{11}\rangle_{A'B'E} 
\eeqa
with the states $\ket{ij}$ on $AB$ and $\Psi_{ij}$ on $A'B'E$. 
Since the key is unbiased and perfectly correlated, we must have 
 $b=c=0$ and $|a|^2=|d|^2=1/2$.
Depending on whether the key is $\ket{00}$ or $\ket{11}$, Eve will hold
the states
\be
\varrho_{0}=Tr_{A'B'}|\Psi_{00}\rangle \langle \Psi_{00}|,\quad
\varrho_{1}=Tr_{A'B'}|\Psi_{11}\rangle \langle \Psi_{11}|
\ee
Perfect security requires $\vr_0=\vr_1$. Thus there exists 
unitaries $U_{00}$ and $U_{11}$ on $A'B'$ such that 
\ben
|\Psi_{00}\rangle&=&\sum_{i}\sqrt{p_{i}}
|U_0\phi_{i}^{A'B'}\rangle|\varphi_{i}^{E}\rangle \nonumber\\
|\Psi_{11}\rangle&=&\sum_{i}\sqrt{p_{i}}
|U_1\phi_{i}^{A'B'}\rangle|\varphi_{i}^{E}\rangle \s.
\een
After tracing out $E$, we will thus get a state of the form Eq.
(\ref{eq:pstate}), where $\vr_{A'B'}=\sum_ip_i|\phi_{i}\>\<\phi_i|$. \blacksquare

It is instructive to see the matrix of a general  $\gamma_1$-state:
\be
\gamma_1=\left[\bea{cccc}
\upl &0&0& \X \\
0& 0&0&0 \\
0&0&0&0\\
\X^\dagger &0&0& \lowr \\
\eea
\right]
\label{eq:block-singlet}
\ee
where the matrix is written in the computational basis on $AB$
i.e. $\ket{00},\ket{01},\ket{10},\ket{11}$ and
the trace norm of block $\X$ is $1/2$. 
Thus $\gamma_1$ looks like a Bell state with blocks instead of c-numbers,
and the condition on $\|\X\|$ can be associated with the fact that Bell states
have the corresponding element (coherence) equal to $1/2$.

Let us briefly sketch the situation where one only demands approximate security for $m=1$. 
Consider in place of $\gamma_1$ an arbitrary state written in similar block form. 
One finds that the condition $||X'||\approx 1/2$,
where $X'$ is the upper right block, is equivalent to the state being close to $\gamma_1$ in norm.  
For the converse direction, one can verify that  in terms of the fidelity 
$F(\varrho_0^E,\varrho^E_1)=\tr|\sqrt\varrho^E_0\sqrt\varrho^E_1|$
\be
||X'||=\sqrt{p_{0}p_{1}}F(\rho^E_0,\rho^E_1)
\ee    
where $p_i$ are probabilities of Alice and Bob to obtain outcome $ii$, and $\rho^E_i$ are the corresponding Eve's states. 
Thus having approximate bit of key, 
i.e. uniformity $p_0\approx p_1\approx 1/2$ and security $F(\rho_1^E,\rho_2^E)\approx 1$ (implying $\rho^E_0\approx \rho^E_1$)
is equivalent to  sharing state close to $\gamma_1$.  The result can be generalized to $m>1$ \cite{huge-key}.

%
and thus that the resulting state be close in norm to some $\pb_1$.

This then completely recasts the drawing of key at a rate $K_D$ 
under local operations and public communication (LOPC)
in terms of distilling $\gamma_m$ states (at a rate of $K_\gamma$ under LOCC).   
Clearly $K_\gamma \leq K_D$ since distilling $\gamma_m$ is a particular way of drawing key.
Additionally, by Theorem 1, any secure protocol which distills $K_D$ is also distilling $\gamma_m$
with $K_\gamma =K_D$ when one considers all of Alice and Bob's lab as the $A'B'$ ancilla.  I.e. if
one applies some protocol coherently (since the original LOPC protocol might be partly classical), 
one distills some $\gamma_m$ at the full rate. We thus have equality of the two rates. 

Before showing that one can have bound entangled states which give secure key,
we provide  examples of both strict and approximate $\pb$ states, which have
an arbitrarily small amount of distillable entanglement 
i.e.  $\K\gg \D$.
 
{\it Example 1.} Consider states
\be
\varrho = p \proj{\psi_+}\otimes \varrho_+ +(1-p)  \proj{\psi_-}\otimes \varrho_-
\label{eq:jstate}
\ee
where $\psi_\pm={1\over \sqrt2} (|00\>\pm|11\>)$ and $\varrho_\pm$
reside on orthogonal subspaces.  One can verify that these
states are particular examples of $\pb_1$, and therefore produce at least one bit
of private key.  
Eve (who holds the purification of the state) can learn the phase of the
state on $AB$, i.e. whether
Alice and Bob hold $\psi_-$ or $\psi_+$.  She can help Alice and Bob  obtain
one singlet by telling them which maximally entangled state they possess. 
Yet she can learn nothing about the
key bit (i.e. whether they have $|00\ra$ or $|11\ra$.  In a sense, Eve can hold
one bit of information but it is the wrong bit of information.  
Such a situation is impossible classically (or with pure quantum states held by Alice and Bob). 

To decrease the distillable entanglement, take $ p=(1+1/d)/2$ and $\vr_\pm$ to be 
two extreme  Werner $d\ot d$ states 
\be
\varrho_s={2\over d^2+d}\Ps,\quad\varrho_a={2\over d^2-d}\Pa
\ee
with $\Pa,\Ps$ the anti/symmetric projectors.
The log-negativity $\Logneg$ which is
an upper bound on the distillable entanglement $\D$ \cite{Vidal-Werner}
amounts in this case to $\Logneg(\varrho)=\log {d+1\over d}$. Thus by
increasing $d$ one can have an arbitrary small amount of distillable
entanglement while keeping one bit of private key.

{\it Example 2.} We take $\vr_\pm$ to be two separable
hiding states $\tau_0$ and $\tau_1$.  We take here those 
given in \cite{ew-hiding}
\begin{equation}
 \tau_0=\vr_s^{\otimes l},
\qquad
 \tau_1= 
 \left[(\vr_a+\vr_s)/2\right]^{\otimes l} \s .
 \label{eq:hiding}
\end{equation} 

By choosing $d$ and $l$ one can make them
arbitrarily indistinguishable under LOCC and arbitrarily orthogonal
(since $\X=(\tau_1-\tau_0)$, orthogonality of the $\tau$'s are needed for security i.e. $\|\X\|$, 
while hiding is needed for low distillability).  
Choosing $p=1/2$, one can show that distilling 
entanglement essentially
reduces to Alice and Bob determining which maximally
entangled state they possess by performing measurements on the hiding state
$\tau$.
Choosing better and better hiding states decreases the
distillable entanglement arbitrarily.  
Again we check this by use of log-negativity; one finds that 
$\Logneg(\varrho)=||\tau_{0}^{\Gamma}-\tau_{1}^{\Gamma}||$ where 
$\Gamma$ stands for partial transpose.  
This quantity has been shown to be an upper bound for 
distinguishability of the hiding states, and 
for suitable choice of $l$ and $d$ it can be 
made arbitrarily small \cite{ew-hiding}.

The idea behind both examples is similar: Alice and Bob share
mixture of two Bell states, with flags which
are flags distinguishable if one has access to the entire state - this gives security,
but are poorly distinguishable by local operations and classical communication, 
which prevents Alice and Bob knowing which Bell state they share,
hence dramatically decreases distillable entanglement.
In both examples however, the states do have nonzero distillable entanglement. For strict 
$\pb$ states, it is not hard to see that they are always distillable. 
It is then clear that any key from bound entangled states can be arbitrarily secure,
but not perfectly secure.

{\it Main result.} We now introduce a bound entangled state which can be shown to have $\K>0$.
We simply take the preceding state, and introduce errors
\be
\rho=\left[\bea{cccc}
{p\over 2}(\tau_0+\tau_1) &0&0&{p\over 2}(\tau_1-\tau_0) \\
0& ({1\over 2} -p )\tau_0&0&0 \\
0&0&({1\over 2} -p)\tau_0& 0\\
{p\over 2}(\tau_1-\tau_0) &0&0& {p\over 2}(\tau_0+\tau_1) \\
\eea
\right]
\label{raw-key}
\ee

One finds that for  
$p\leq 1/3$ and ${}^l\!\!\sqrt{1-p\over p}(d-1)\geq d$ 
the state has positive partial transpose (PPT)  being therefore bound entangled 
\cite{bound}.

Now, we take $n$ copies, and apply the recurrence distillation protocol of \cite{BBPSSW1996} 
without the twirling step. The resulting state is 
\be
\hskip-3mm \rho'=
{1\over {N}}\left[\bea{cccc}
[{p\over 2}(\tau_0+\tau_1)]^{\otimes n} &0&0&\hskip-4mm[{p\over
  2}(\tau_1-\tau_0)]^{\otimes n}\\
0& \hskip-4mm[({1\over 2}-p) \tau_0]^{\otimes n}&0&0 \\
0&0 &\hskip-4mm[({1\over 2}-p) \tau_0]^{\otimes n}&0 \\
{}[{p\over 2}(\tau_1-\tau_0)]^{\otimes n} &0&0& 
\hskip-4mm[{p\over 2}(\tau_0+\tau_1)]^{\otimes n}\\
\eea
\hskip-2mm\right]
\label{rhoprim}
\ee
where ${N}= 2p^{n}+2\left(1/2-p\right)^{n}$.
To see that Alice and Bob have arbitrarily secure key, we check 
that the trace  norm 
of off-diagonal block $\|\X\|$ tends to $1/2$:
\be
\bigl\|[{p\over 2}(\tau_1-\tau_2)]^{\otimes n}/N\bigr\|={1\over 2} 
\bigl(1-{1\over 2^l}\bigr)^n {1\over 1+ \bigl({1-2p\over 2p}\bigr)^n}
\ee
Now, for $p> 1/4$ the norm can be arbitrarily close to $1/2$ if 
we had previously taken $l$ large
enough, and now take large $n$.
Given such $l$, one could always have initially chosen $d$ 
to satisfy the PPT condition of the initial state 
(\ref{raw-key}), so that the state $\rho'$ is PPT 
(as it is obtained from $\rho$ by LOCC). 

{\it Remark.} Note that we need to use large $l$ for security, large 
$n$ for the state to approximate perfect key, and large $d$ 
for the state to be PPT. Indeed, large $d$ is needed for $\tau_i$
to be hiding states, and if they are not hiding, then the state 
would be distillable by distinguishing between them, and then distilling
the correlated singlet.  

Thus we have shown that we can get arbitrarily secure  bit from bound entangled states
The structure of our states sheds perhaps for the first 
time some light 
on the phenomenon of bound entanglement: they can contain singlets that are so ``twisted'',
they cannot be distilled, but they can exhibit their quantum character 
through privacy.  
This explanation probably cannot be applied to low-dimensional bound entangled states.

Having show that one can draw one bit of key, we now
show that Alice and Bob can draw key at a nonzero asymptotic rate, using 

{\it Lemma 1.
For any state $\psi_{ABA'BE}$ consider the state $\vr_{ABE}$ 
emerging after measurement on $AB$ in the standard basis. 
The latter state does not change under twisting.}
(the proof boils down 
to direct checking) 

Since trace norm of the off-diagonal block (\ref{rhoprim}) 
of the state is close to $1/2$, by use of polar decomposition,
one finds twisting operation after which {\it trace} of the block $\X$ 
is equal to its trace norm. For such new state $\rho''$, by Lemma 1,
Eve's states correlated with outcomes of $AB$ 
measurements are still the same as for $\rho'$. Now however,
after  tracing out $A'B'$, the state is close to singlet. Clearly, 
the problem  is reduced to drawing key from outcomes of measurement,
from a state close to singlet, which can be done, for example, by the 
protocol of  Devetak and Winter \cite{DevetakWinter-hash}.  
%
As we have already noted, this will draw $\gamma$ states at the same rate as $K_D$ when
the corresponding classical protocol is applied coherently.
%


We now provide a general upper bound on $\K$ in terms of the relative
entropy of entanglement
$\E(\rho):=inf_{\sigma\in\sep}S(\rho||\sigma)$, with
$S(\rho||\sigma):=\tr[\rho(\ln{\rho}-\ln\sigma)]$ and $\sep$ being the
set of separable states.  Namely we have


{\it Theorem 2.
  $\K(\rab)\leq\E^\infty(\rab)$.  
where $\E^\infty$ is the regularization of the relative entropy of entanglement
$\E^\infty(\rho):=\lim_{n\rightarrow\infty}\E(\rho^{\ot n})/n$.}

Our proof is inspired by the idea that 
transition rates are bounded by LOCC monotones \cite{Michal2001}, 
yet it needs essentially new techniques, mostly due to 
possibility of large scaling of the size of the ancilla $A'B'$ 
with the number of obtained bits of key. We present it in \cite{keyhuge}.

Since we can have $\E(\rho)<E_c(\rho)$ the above theorem implies that for some
states, the key rate will be strictly less than the entanglement cost, and
in fact, can be made arbitrarily small for fixed $E_c$.  E.g.
for  anti-symmetric Werner state $\vr_{a}$
we have  $E_c(\vr_a)=1$ \cite{MatsumotoY03} while $\erinf(\vr_{a})=\log{(d+2)/d}$
which can be arbitrarily low. 


In summary, we have found that in general $\D\leq\K\leq\erinf\leq E_c$ with strict inequalities
$\D<\K<E_c$ and $\D<\erinf$ also possible (the latter was shown 
previously in \cite{AVC-2003}; our result allows for easy construction of 
new examples).  One can even have $\K>0$ for bound
entangled states. This implies that the rate of
distillable key is not only an operational measure of entanglement, but
is also non-trivial in that it is not equal to 
other known operational measures: $E_c$ and $\D$. 
This is also likely to be true for the {\it quantum key cost} $K_c$ which we define to
be the minimum size $m$ of $\pb_m$ required to form a state in the
asymptotic limit.   These results also put into question the
possibility of ``bound information'' for bipartite systems conjectured in 
\cite{boundinfo-gw}, although the phenomena may well exist 
for distributions
derived from other bound
entangled states.
Our results also suggest 
that the qualitative equivalence between
privacy and distillability in $2\otimes2$ \cite{acin-gisin} is likely
to be due to the fact that in low dimensions, bound entanglement does not
exist. 

One could define a unit of privacy,  by calling  $\pb_1$ {\it irreducible}, 
if one and only one bit of privacy can be obtained from it. 
Irreducible private state  may therefore be thought of as the basic unit state
of privacy, much as the singlet is the basic unit of entanglement theory
(although not all $\gamma$ states are equivalent to each other, thus one thinks 
of $\gamma_m$ in its entirety).
>From theorem 2 it follows that irreducibility 
can be imposed by demanding that $\pb_1$ have a relative entropy of
entanglement of one. However we do not know if this condition is too strong.

Here our interest in privacy is motivated by 
the fundamental insight it gives into entanglement -- there seems 
to exist a deep connection between the entanglement cost of PPT
states, and privacy.  In terms of cryptographic protocols,
the states considered here can be incorporated into an actual scheme
by performing a suitably randomized tomography protocol 
on the obtained states to verify that they are
indeed close to the expected form.  Such a protocol is highly
inefficient, but appears to be secure for binding entanglement channels,
although the scaling of security parameters may be qualitatively
different than in BB84.  
Determining how efficient such a protocol
could be is an interesting open problem. 

\begin{acknowledgments}
This work is supported by EU grants RESQ (IST-2001-37559),
QUPRODIS (IST-2001-38877) and
PROSECCO (IST-2001-39227), and JO additionally
acknowledges 129/00-1 of the ISF, and a grant from the Cambridge-MIT Institute.
We thank A. Acin, R. Horodecki, A. Winter and N. L\"utkenhaus for 
helpful feedback and T. Eggeling for sending us an advance copy of
the sequel to Ref. \cite{ew-hiding}.
\end{acknowledgments}

\bibliography{../refjono,../refmich}

\end{document}